\documentclass[a4paper,12pt]{article}
\usepackage{amsmath}
\usepackage{graphics}
\usepackage{graphicx}
\usepackage{latexsym}
\usepackage{float}
\usepackage{latexsym}
\usepackage{amsfonts}
\usepackage[mathscr]{eucal} 
\usepackage{amssymb}
\usepackage{epsf}
\usepackage{psfrag}
\usepackage{pstricks}
\usepackage{afterpage}
\usepackage{indentfirst}
\usepackage{multido}
\usepackage{array}
\usepackage{tabularx}
\usepackage{changebar}
\allowdisplaybreaks
\newcommand{\rgu}{\mathring{u}}
\newcommand{\rgv}{\mathring{v}}

\newcommand{\ZZs}{\begin{changebar}}
\newcommand{\ZZe}{\end{changebar}}
\date{}
\begin{document}
\title{\bf{Some asymptotic results for bifurcations in \\ 
           pure bending of rubber blocks}}
\author{Ciprian D. Coman,
        Michel Destrade}
        \date{2008}
        
\maketitle
%
\begin{abstract}
\noindent The bifurcation of an incompressible neo-Hookean thick hyperelastic plate
with a ratio of thickness to length $\eta$ and subject to pure bending is
considered within a plane-strain framework. 
The two incremental equilibrium equations corresponding to a nonlinear
pre-buckling state of strain are
reduced to a fourth-order linear eigenproblem that displays a multiple
turning point. It is found that for $0<\eta<\infty$ the plate experience an Euler-type
buckling instability which in the limit $\eta\to\infty$ degenerates into
a surface instability. Singular perturbation methods enable us to capture 
this transition, while direct numerical simulations corroborate the analytical
results.\\[0.3cm]
{}\\
{\bf{Keywords:}} hyperelastic plates, incremental equations, turning points, boundary layers.
\end{abstract}
%
\newpage
\section{Introduction}
%
%
\noindent The development of compressive stresses in mechanical structures is well known
to be responsible for Euler-type buckling instabilities. What is less
recognised is that such scenarios are likely to occur in a number of cases
that, apparently, are of a completing different nature. A typical example is
the phenomenon of stress concentration in perforated thin elastic plates
subjected to tension. Usually, the holes act as stress concentrators that can
be completely or only partially surrounded by compressed regions. If the
pulling forces are sufficiently strong an out-of-plane bending instability
is experienced locally near the sites of the holes. A systematic investigation
of problems of this nature has recently been initiated by Coman {\emph{et al.}}
\cite{cdc_dmh_1,cdc_apb:2007a,cdc_apb:2007b,cdc_apb:2007c}.\par
A second example where Euler-type buckling is indirectly encountered is
provided by the pure bending of a thin and short elastic tube. The curved
configuration adopted by the tube is
characterised by compressive axial stresses on the concave side, whereas
tension will prevail on the convex part. Experience shows that a regular
instability pattern consisting of many little ripples will develop along the
former region, eventually leading to the creation of one or several kinks
that signal the collapse of the tube.\par
The stability problem of pure bending in thin-walled tubular structures has a long
history and there is a vast mechanical engineering literature dealing with
various aspects; some of it is aptly summarised in the authoritative 
account of Kyriakides \& Corona \cite{stelios:07}. On the mathematical side,
noteworthy contributions in the present context are the works by 
Seide \& Weingarten \cite{seide:61} and those by Tovstik {\emph{et al.}}
(briefly summarised in \cite{tovstik:00}). The former investigation is based
on the Donnell-von K\'{a}rm\'{a}n buckling equations linearised around a
variable-coefficient membrane state of stress. The resulting boundary value
problem was analysed numerically with the help of the Galerkin method, and 
it was found that the circumferential shape of the buckled cylinder displays
a small dimple on the compressed side.
Several versions of the same problem have been re-considered in \cite{tovstik:00}
from the point of view of asymptotic analysis. Both works just now mentioned 
made the simplifying assumption that the rippling pattern is the same at every 
point along the axis of the cylinder or, in other words, a solution with
separable variables was {\emph{a priori}} postulated. The assumption is
sensible for short tubes (which are fairly stiff), but it is
inadequate for modelling the collapse in the elasto-plastic regime for
moderate lengths, which turns out to require a very different 
approach (cf.~\cite{stelios:07}).\par
\begin{figure}[!h] 
\centerline{\input{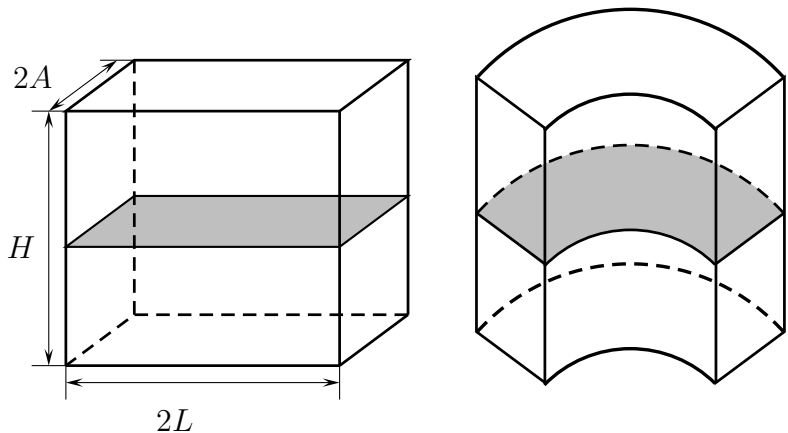}}
\center{\parbox[b]{5in}{
\caption{\label{pic_gen}
    {\small{Cylindrical bending of a rubber block; left: {\emph{reference}}
    configuration, right: {\emph{current}} configuration. Shaded areas indicate
    two generic cross-sections perpendicular to the vertical axis.}}}}}
\end{figure}

Our main aim in the present investigation is to re-visit the pure bending of a rubber
block deforming in plane strain, a problem that has several points in common 
with the tube bending mentioned above. Unfortunately, the literature in this
area has focused mainly on describing the deformation itself rather than its 
potential bifurcations.
The typical scenario is outlined in Figure~\ref{pic_gen}: the undeformed
configuration is shown in the left-hand sketch and is characterised by the
geometric parameters $2L$ ({\emph{length}}), $H$ ({\emph{height}}), and $2A$
({\emph{thickness}}); the deformed block appears on the right in the same Figure. The
plane-strain hypothesis simplifies the problem considerably, since one needs 
deal only with cross-sections (shown shaded) perpendicular to the vertical
axis of the block and situated sufficiently far away from the lower and upper faces.\par
%
%
Pioneering work on pure bending stability was carried out by Triantafyllidis \cite{tria:80}, who examined
incremental bifurcation equations for a couple of piecewise power law constitutive models,
including a hypoelastic one. He pointed out that the underlying instability
mechanism is a surface instability similar to that encountered in the
plane-strain half-space problems discussed by Hill \& Hutchinson \cite{hill:75}
or Young \cite{young:76}. Haughton \cite{dmh:99} performed a similar analysis for
hyperelastic materials in a three-dimensional context (neo-Hookean, mostly)
and allowed for vertical compression as well. The instability was found to be of Euler-type but his
interpretation of some of the results is wrong (as we shall explain in \S3). 
A novel feature in \cite{dmh:99} is the interaction between two different 
modes of instability, one due to pure bending, the other related to compression. 
Dryburgh \& Ogden \cite{rwo:99} introduced thin coatings on the curved boundaries of the
bent block and made comparisons with the uncoated case. Their findings show
that, relative to the latter case, bifurcation is generally promoted by the
presence of surface coating, on either or both curved boundaries, that is the
bifurcation occurs at smaller strains. The relative sizes of the shear moduli
for the coating and, respectively the bulk material was found to play an important
role in describing this phenomenon.\par
The finite elasticity works reviewed above share a common feature in that they
all deal with incompressible materials. So far little is known about the role
played by compressibility on the bifurcation behaviour in pure bending.
The reason might be rooted in the absence from the literature of a manageable 
closed-form expression for the pre-bifurcation deformation. 
Aron \& Wang \cite{aron:95a,aron:95b} touched upon issues like existence
and uniqueness for bending deformations in unconstrained elastic materials,
while Timme {\emph{et al.}} \cite{otto:02} used Hencky's compressible
elasticity model to investigate closed-form solutions for cylindrical bending.
They succeeded in deriving explicit expressions for the bending
angle and moment in terms of the circumferential stretches on the curved
boundaries. The solution is quite involved and it seems unlikely to be useful
for anything but numerical calculations. Moreover, the particular Hencky elasticity
framework is restricted by moderate deformations only.\par
A critique of bifurcation phenomena in pure bending was given by Gent \&
Cho \cite{gent:99} who pointed out that their experiments did not agree with 
the theoretical predictions based on the surface-instability concept proposed
in \cite{tria:80}. In particular, they found that the instability occurs for a smaller 
degree of compression and the block adopts a configuration with a 
small number of {\emph{sharp creases}} on the inner surface. To fully explain this
observation would require a nonlinear {\emph{post-buckling}}
analysis because the bifurcation involved is probably of {\emph{subcritical}} type. 
We note in passing that Gent \& Cho's
creases are, to a certain extent, very similar to those encountered on the
curved surface of severely torsioned stocky rubber cylinders \cite{penn:1976}.
These phenomena are likely to be related to the failure at the boundary of the 
{\emph{complementing condition}} (see \cite{mac:05} and the reference
therein), and they fall outside the scope of our study.\par
With this background in mind, we shall re-consider in the next sections a
particular instance of the pure bending problems taken up in
\cite{dmh:99,rwo:99}. The aim is to elucidate the nature of the instability
and to analyse the mathematical structure of the governing boundary value problem 
when $\eta\equiv{A/L}\gg{1}$. To avoid ``missing the forest for the trees'',
the model investigated will be confined to incompressible neo-Hookean 
materials. In \S2 these assumptions are showed to yield an 
eigenproblem for a fourth-order partial differential equation with
variable coefficients, subsequently simplified by seeking a solution with
separable variables. Direct numerical simulations reveal an Euler-type
buckling phenomenon for $0<\eta<\infty$, but in the limit $\eta\to\infty$ this
degenerates into a surface instability. Some erroneous interpretations
proposed by previous investigators (e.g., \cite{tria:80} or \cite{dmh:99}) are 
also corrected here for the first time. As demonstrated in \S4, the transition
regime between the two different forms of instabilities can be efficiently
captured by singular perturbation methods. The two contrasting asymptotic
methods employed are discussed separately in \S\ref{wkb_approach} (WKB) and,
respectively, \S\ref{blt} (boundary layers). The former would seem to be the
most appropriate because the differential equation in question has variable coefficients. 
However, it transpires that a conventional boundary-layer
analysis sheds more light and helps us to steer clear from the 
confusion created by the presence of a {\emph{multiple turning point}}. Unlike in the
recent studies \cite{cdc_dmh_1,cdc_apb:2007a,cdc_apb:2007b,cdc_apb:2007c},
here turning points play no role whatsoever (the same is true for the related
works \cite{fu:02a,dmh_chen:02}).
The paper concludes with a discussion of the results obtained, together with
suggestions for further study.
%
%
\section{Overview of the model}
%
%
\noindent Finite pure bending of incompressible hyperelastic
materials is discussed in a number of books like, for example, \cite{green:60}
or \cite{rwo:97}. To make the paper reasonably self-contained, we summarise
some of those ideas below.\par
\begin{figure}[!h] 
\centerline{\input{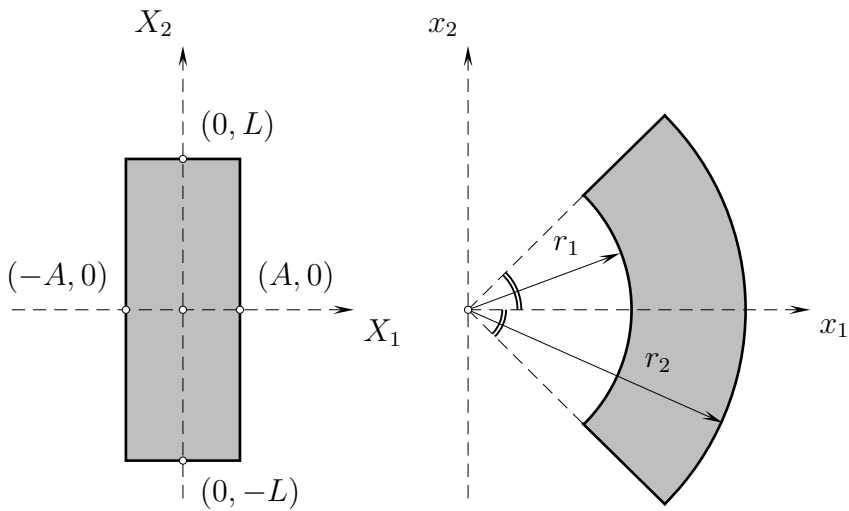}}
\center{\parbox[b]{5in}{
\caption{\label{pic_part}
    {\small{The {\emph{undeformed}} (left) and {\emph{deformed}} (right) cross-sections of the rubber
            block shown in Figure~\ref{pic_gen}. Bending is symmetric with
            respect to the $x_1$-axis so that the two angles marked are
            congruent and equal to $\omega_0$; see the text for more details.}}}}}
\end{figure}

The reference configuration of the initially undeformed rectangular
cross-section of the hyperelastic block is the region (see Figure \ref{pic_part})
$$
\mathcal{B}_R:=\left\{(X_1,X_2)\in\mathbb{R}^2\;\big{|}\;-A\leq{X_1}\leq{A},\; -L\leq{X_2}\leq{L}\right\}\,.
$$
Supposing that the block is bent (symmetrically with respect to the $x_1$-axis)
into a sector of circular cylindrical tube, the current configuration 
of the deformed cross-section is easily represented in polar co-ordinates by the domain
$$
\mathcal{B}_C:=\left\{(r,\theta)\in\mathbb{R}\times(0,2\pi]\;\big{|}\;-r_1\leq{r}\leq{r_2},\; 
                                 -\omega_0\leq{\theta}\leq{\omega_0}\right\}\,.
$$
%
%
Rivlin \cite{rivlin:49} showed that for an incompressible elastic
material this type of deformation may be described by the mapping
\begin{equation}
\label{prebuck_map}
r = (d+2X_1/\omega)^{1/2},\qquad \theta = \omega{X_2}\,,
\end{equation}
where $d$ is a quantity determined by the particular constitutive law
adopted and $\omega$ can serve as a control parameter as it is related to the {\emph{angle of bending}},
$\omega_0:=\omega{L}$. Since the plate cannot be bent into itself, we should
require that
\begin{equation}
\label{constraint:1}
0<\omega_0\leq\pi\,,
\end{equation}
an assumption used tacitly henceforth.
Although the deformation recorded in (\ref{prebuck_map}) seems to have an
iconic status among workers in elasticity, it is clear that the kinematics
afforded by that expression are somewhat restricted. The lines $X_1=\text{const.}$
become arcs of the circle $r=\text{const.}$, while the lines $X_1=\text{const.}$ are
transformed in lines $\theta=\text{const.}$; in other words, ``cross-sections''
perpendicular to the vertical symmetry axis of $\mathcal{B}_R$, remain orthogonal
to the deformed axis of the current configuration $\mathcal{B}_C$. This is somewhat at
odds with the commonly accepted point of view in structural mechanics,
according to which pure bending of thick sandwich panels (e.g.,
\cite{allen:69}) is based on models that allow cross-sections to slide relative 
to the normal to the deformed axis. Nonetheless, the {\emph{nonlinear}} mapping
(\ref{prebuck_map}) is still a sensible choice for the type of questions we want to
answer, at least in a first approximation. \par
The bifurcation analysis carried out in this work is based upon
linearising the plane-strain equations of finite elasticity around
the {\emph{nonlinear}} pre-buckling deformation (\ref{prebuck_map}). This approach to
linearised or incremental bifurcations is well established
(see \cite{biot:65,rwo:97}, for example), so we shall not rehearse it
here. Instead, we limit ourselves to pointing out the key steps that lead to 
our eigenproblem.\par
We shall assume that the constitutive behaviour of the material
is characterised by a strain-energy function $W\equiv W(\lambda_r,\lambda_\theta)$, where the principal stretches
$\lambda_r$ and $\lambda_{\theta}$ are associated with the Eulerian principal
directions $\mathbf{e}_r$ and, respectively, $\mathbf{e}_{\theta}$.
Due to the incompressibility constraint these can be written as
$$
\lambda_r = \lambda^{-1}\qquad\text{and}\qquad \lambda_{\theta}\equiv\lambda:=\omega{r}\,,
$$
which defines the notation $\lambda$.
According to \cite{rwo:97},
the two-dimensional version of the incremental 
equations of equilibrium for incompressible elasticity read
\begin{equation}  
\label{increment}
\text{div}\mathring{\mathbf{s}} = 0,\qquad \text{div} \mathring{\mathbf{u}} = 0\,,
\end{equation}
where $\mathring{\mathbf{u}}=(\mathring{u}(r,\theta),\mathring{v}(r,\theta))$ is the incremental displacement
field, and  $\mathring{\mathbf{s}}$ denotes the incremental nominal stress tensor
with components
$$
\mathring{s}_{ij} = L_{ijkl}\mathring{F}_{kl}+p\mathring{F}_{ij}-\mathring{p}\delta_{ij}\,,\quad\qquad
i,\,j\in\{r,\,\theta\}\,.
$$
Here $\mathring{p}$ is the increment in the Lagrange multiplier 
$p\equiv p(r,\theta)$ (the ``hydrostatic pressure''), while $\mathring{F}_{ij}$ represent
the components of the incremental deformation gradient,
$$
\mathring{\mathbf{F}} =\begin{bmatrix}
                       \rgu,_{r} & \frac{\displaystyle 1}{\displaystyle r}(\rgu,_{\theta}-\rgv)\\[0.2cm]
                       \rgv,_{r} & \frac{\displaystyle 1}{\displaystyle r}(\rgu + \rgv,_{\theta}) 
                       \end{bmatrix}\,.
$$
Finally, $L_{ijkl}$ are the components of the fourth-order tensor of
instantaneous incremental moduli which,
in Eulerian principal axes has $15$ 
independent non-zero such components (cf.~\cite{rwo:97})
\begin{subequations}
\label{moduli}
\begin{alignat}{2}
\label{moduli:1}
L_{iijj} &=\lambda_i\lambda_jW_{ij}\,, \\[0.2cm]
\label{moduli:2}
L_{ijij} &=\begin{cases}
            \frac{\displaystyle \lambda_i^2(\lambda_iW_i-\lambda_jW_j)}
                 {\displaystyle \lambda_i^2-\lambda_j^2} \qquad &\text{if}\quad 
                                       {i}\neq{j},\;\lambda_i\neq\lambda_j\,,\\[0.2cm]              
            \frac{\displaystyle 1}{\displaystyle 2}(L_{iiii}-L_{iijj}+\lambda_iW_i)\qquad 
           &\text{if}\quad i\neq{j},\;\lambda_i=\lambda_j\,,
           \end{cases}\\[0.2cm]
\label{moduli:3}
L_{ijji} &= L_{jiij}=L_{ijij}-\lambda_iW_i\,,
\end{alignat}
\end{subequations}
with $W_i\equiv\partial{W}/\partial\lambda_i$, $W_{ij}\equiv\partial^2{W}/\partial\lambda_i\lambda_j$,
and the summation convention does not apply. \par
Direct calculations show that the system of equations (\ref{increment}) can be reduced to
{\setlength{\multlinegap}{50pt}
\begin{multline} 
\label{beef_pde_1}
r^2\mathring{p,}_{r} = r\left[r(L_{1111}' - L_{1122}' + p,_{r}) + L_{1111} + L_{2222} - 2L_{1122}\right]\rgu,_{r}\\
                   +{r^2}(L_{1111} - L_{1122})\rgu,_{rr} + L_{2121}(\rgu,_{\theta\theta} - \rgv,_{\theta}) +
                   {r}L_{2112}\rgv,_{r\theta}\,,
\end{multline}}
\vspace*{-0.5cm}
{\setlength{\multlinegap}{70pt}
\begin{multline}
\label{beef_pde_2}
r\mathring{p,}_{\theta} = ({r}L_{1212}' + L_{1212})(r\rgv,_{r}+\rgu,_{\theta}-\rgv) + {r^2}L_{1212}\rgv,_{rr}\\
                   + {r}(L_{2112} + L_{1122} - L_{2222})\rgu,_{r\theta}\,.
\end{multline}}
\hspace*{-0.2cm}To avoid overdoing the notation we have used the
correspondence $r\to{1}$ and $\theta\to{2}$ for the incremental moduli, and
have indicated their derivatives with respect to $r$ by dashes.
A further simplification is afforded by the incompressibility condition which
allows us to deduce the existence of a potential $\phi\equiv\phi(r,\theta)$ such that
\begin{equation}
\label{potents}
\mathring{u} = \frac{1}{r}\frac{\partial\phi}{\partial\theta},\qquad\quad
\mathring{v} = -\frac{\partial\phi}{\partial r}\,.
\end{equation}
The upshot of this observation is that the two equations
(\ref{beef_pde_1},\ref{beef_pde_2}) can now be combined into a single
partial differential equation for the potential function. After some routine (but
lengthy) manipulations, we end up with 
\begin{equation}
\label{beef_potent}
\sum_{j=1}^{4}\mathcal{L}_j[\phi]=0\,,
\end{equation}  
with $\mathcal{L}_j$ partial differential operator of the $j$-th order defined
according to
\begin{alignat*}{1}
\mathcal{L}_4 &:=\alpha\,{r^4}\frac{\partial^4}{\partial r^4} + 
                     2\beta{r^2}\frac{\partial^4}{\partial r^2\partial\theta^2}+
                     \gamma\frac{\partial^4}{\partial\theta^4}\,,\\[0.2cm]
\mathcal{L}_3 &:= 2r^3(r\alpha)'\frac{\partial^3}{\partial r^3} + 
                  2r^3\left(\frac{\beta}{r}\right)'\frac{\partial^3}{\partial{r}\partial\theta^2}\,,\\[0.2cm]
\mathcal{L}_2 &:= r^4\left[\alpha''+\left(\frac{\alpha}{r}\right)'\right]\frac{\partial^2}{\partial{r}^2}-
            r^2\left[\alpha''+\left(\frac{\alpha+2\beta}{r}\right)'-\frac{\gamma}{r^2}\right]
                        \frac{\partial^2}{\partial\theta^2}\,,\\[0.2cm]
\mathcal{L}_1 &:= -r^3\left[\alpha''+\left(\frac{\alpha}{r}\right)'\right]\frac{\partial}{\partial{r}}\,,
\end{alignat*}
and 
\begin{equation*}
\alpha(r):=L_{1212}\,,\qquad
\gamma(r):=L_{2121}\,,\qquad
\beta(r):= \frac{1}{2}(L_{1111}+L_{2222})-(L_{1122}+L_{2112})\,.
\end{equation*}
The form (\ref{beef_potent}) of the bifurcation equation is valid for any
choice of incompressible hyperelastic material but, as it stands, the model
is not easily amenable to analytical work. Before further simplifications are
implemented, we must address the issue of boundary conditions for
(\ref{beef_potent}).\par
The two curved boundaries of $\mathcal{B}_C$ are taken to be traction-free, a
constraint which demands
\begin{subequations}
\label{bc_potent}
\begin{alignat}{2}
\label{bc_potent_1}
&\alpha{r^3}\phi,_{rrr}-(2\beta+\alpha)(\phi,_{\theta\theta}-r\phi,_{r\theta\theta})=0\,,\qquad
                &&\text{for}\quad (r,\theta)\in \{r_1,r_2\}\times(0,2\pi]\,,\\[0.2cm]
\label{bc_potent_2}
&\phi,_{\theta\theta}+r\phi,_{r}-{r^2}\phi,_{rr} = 0\,,\qquad
                &&\text{for}\quad (r,\theta)\in \{r_1,r_2\}\times(0,2\pi]\,.
\end{alignat} 
\end{subequations}
These conditions can be obtained as a particular case of the calculations of
Dryburgh \& Ogden \cite{rwo:99}, to which the reader is referred for more
information.\par
Next, we look for separable solutions of the bifurcation equation (\ref{beef_potent}) in
the form
\begin{equation}
\label{normal_mode}
\phi(r,\theta) = \Phi(r)\cos(m\theta)\,,
\end{equation}
where $m\in\mathbb{N}$ is the azimuthal mode number related to the
number of ripples on the compressed side of the rubber block, while $\Phi(r)$ 
is the infinitesimal amplitude of this cosine rippling pattern. Several types
of boundary conditions are possible for the straight boundaries of
$\mathcal{B}_C$. Following \cite{dmh:99,rwo:99} we consider zero incremental
displacement in the radial direction and vanishing normal traction. It can be
shown that these conditions are satisfied as long as 
\begin{equation}
\label{mode_number}
m = \frac{n\pi}{\omega{L}}\,,
\end{equation}  
for some positive $n\in\mathbb{Z}$. With this information in hand, all
that remains to be done is to choose a constitutive model and carry out the
simplification of (\ref{beef_potent}) with the help of the assumed form
solution recorded in (\ref{normal_mode}).\par
The bulk material is modelled by a simple neo-Hookean strain-energy function
specialised to plane-strain elasticity,
$$
W(\lambda_r,\lambda_{\theta}) = \frac{1}{2}\tau(\lambda_r^2+\lambda_{\theta}^2-2),
$$
$\tau$ being the ground-state shear modulus of the material.
As shown by Rivlin \cite{rivlin:49} and subsequently discussed by 
others \cite{tria:80,dmh:99,rwo:99}, for this particular choice
of constitutive law the constant $d$ in (\ref{prebuck_map}) is determined by
$$
d = \frac{L^2}{\omega_0^2}(1+4\eta^2\omega_0^2)^{1/2}\,.
$$
On making use of (\ref{normal_mode}) in (\ref{beef_potent}) results in an ordinary
differential equation which, when expressed in the non-dimensional 
variables
$$
\rho:=\frac{r}{L}\,,\qquad \mu:=n\pi\,,\qquad \eta:=\frac{A}{L}\,,
$$
can be cast as 
\begin{equation}
\label{beef_ode}
\Phi''''+\mathcal{P}(\rho)\Phi'''+\mathcal{Q}(\rho)\Phi'' +
         \mathcal{R}(\rho)\Phi' + \mathcal{S}(\rho)\Phi = 0,\quad{\rm on}\quad \rho_1 < \rho < \rho_2\,.
\end{equation}
Above, the dashes denote derivatives with respect to $\rho$ and 
\begin{alignat*}{2}
&\mathcal{P}(\rho) :=-\frac{2}{\rho}\,,\quad
&&\mathcal{Q}(\rho) := \left[\frac{3}{\rho^2}-
                            \mu^2\left(\omega_0^2\rho^2+\frac{1}{\omega_0^2\rho^2}\right)\right],\nonumber\\
&\mathcal{R}(\rho) :=-\frac{1}{\rho}\left[\frac{3}{\rho^2}+
                   \mu^2\left(\omega_0^2\rho^2-\frac{3}{\omega_0^2\rho^2}\right)\right],\quad
&&\mathcal{S}(\rho) :=\mu^4\,.
\end{alignat*}
Note that the inner and, respectively, the outer curved surfaces of the current
configuration become 
\begin{equation}
\label{boundaries}
\rho_{1,2} = \frac{1}{\omega_0}\left[(1+4\eta^2\omega_0^2)^{1/2} \pm 2\eta\omega_0\right]^{1/2}\,,
\end{equation}
while the principal stretch in the $\mathbf{e}_{\theta}$-direction assumes the simple form
\begin{equation}
\label{r_stretch}
\lambda = \omega_0\rho\,.
\end{equation}
The solution of (\ref{beef_ode}) is found subject to the 
non-dimensional boundary conditions obtained from (\ref{bc_potent}) via (\ref{normal_mode}),
\begin{subequations}
\label{bc_ode}
\begin{alignat}{2}
\label{bc_ode:1}
\Phi''' - \mu^2\left(\omega_0^2\rho^2+\frac{2}{\omega_0^2\rho^2}\right)\Phi' 
        + \frac{\mu^2}{\rho}\left(\omega_0^2\rho^2+\frac{2}{\omega_0^2\rho^2}\right)\Phi &= 0\,,\qquad
           &&\text{for}\quad\rho = \rho_{1,2}\,,\\[0.2cm]
\label{bc_ode:2}
\Phi'' - \frac{1}{\rho}\Phi'+\frac{\mu^2}{\omega_0^2\rho^2}\Phi &= 0\,,\qquad
           &&\text{for}\quad\rho = \rho_{1,2}\,.
\end{alignat}
\end{subequations}
The normal-mode approach has reduced the bifurcation analysis to the study of
a standard ordinary eigenproblem for $\Phi(\rho)$ and $\omega_0\in(0,\pi)$. While for
structural mechanics problems (e.g., \cite{alfutov:00}) this route is free of
pitfalls, in finite elasticity it is only deceptively so. The danger is that
the bent block might develop shear bands or other material instabilities
before the compressed inner surface starts to wrinkle. Such occurrences are
heralded by a loss of ellipticity in the partial differential 
equation (\ref{beef_potent}); unfortunately, they remain undetected by
(\ref{beef_ode}). Conveniently, the use of a neo-Hookean
constitutive law precludes any form of material instabilities (note that
$\mathcal{L}_4$ is strongly elliptic in this case). Such exotic effects, however,
were accounted for in \cite{tria:80}, but it was found that
the surface instability was always the first to occur.     
%
%
%
%
%
\section{Numerical experiments}
\label{numerics}

%
\noindent The stability of the bent rubber block is now investigated numerically, the
starting point being the eigenproblem (\ref{beef_ode},\ref{bc_ode}) formulate
in \S2. Our first objective is to find out the dependence of the critical bending angle
$\omega_0$ in terms of the aspect ratio $\eta\equiv{A/L}$. It is
expected that an Euler-type buckling instability is experienced for
$0<\eta<\infty$, but in the limit $\eta\to\infty$ this behaviour degenerates
into a surface instability. Such behaviour is consistent with the results of 
Dryburgh \& Ogden \cite{rwo:99} and Haughton \cite{dmh:99}, although it
appears that earlier investigators \cite{tria:80} reported only surface
instabilities for some other choices of constitutive behaviour. Unfortunately,
a direct comparison with those results is not possible here but, intuitively, one
would expect that the finite thickness of the block should set a length-scale
for the instability pattern.\par
\begin{figure}[!h]
    \unitlength=1cm
      \psfragscanon
        \centerline{\includegraphics[width=10cm,height=10.0cm]{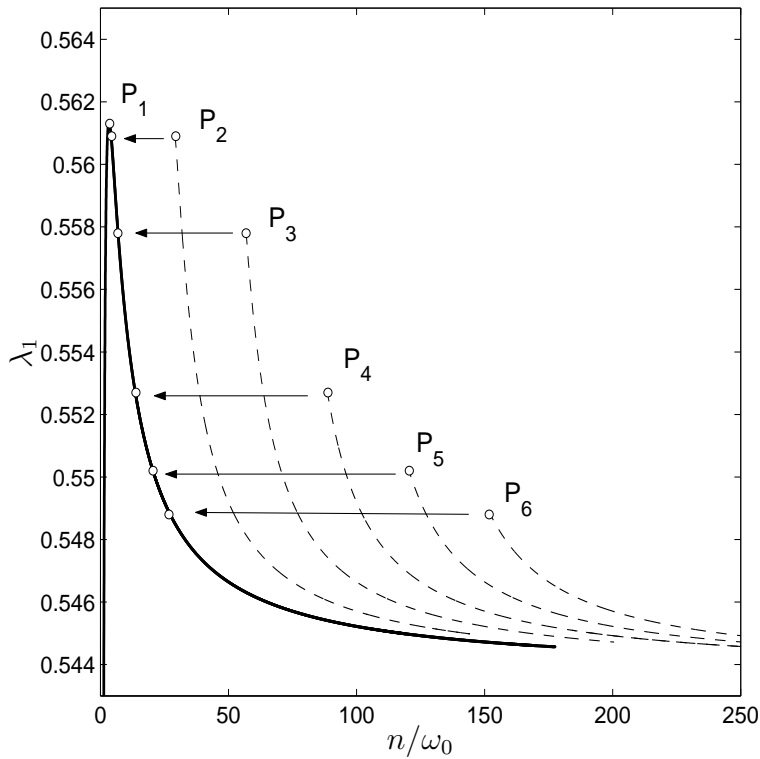}}
        \begin{center}
       \parbox[b]{14cm}
          {\caption[]
           {\small{Plot of the critical stretch
            $\lambda_1\equiv\lambda(\rho_1)$ against the mode number $m\equiv{n}/\omega_0$,
            as obtained by direct numerical integration of the eigenproblem 
            (\ref{beef_ode},\ref{bc_ode}) for a sample of aspect ratios
            $\eta$. The maximum principal stretch for each individual case
            considered is marked by a small circle and corresponds to the points:
            $P_1$~($\eta=1.0$), $P_2$~($\eta=3.0$), 
            $P_3$~($\eta=5.0$), $P_4$~($\eta=10.0$), $P_5$~($\eta=15.0$), 
            and $P_6$~($\eta=20.0$). These maxima are attained
            for $n=1$, except for $P_1$ which corresponds to $n\approx{2.43}$.           
          }}\label{n_curves}}
        \end{center}
\end{figure}

A rather peculiar feature of our eigenproblem is the
dependence of the mode number $m$ in (\ref{mode_number}) on the bending angle.
In order to identify the former quantity for a given $\eta$, we shall plot the principal stretch
on the curved inner boundary $\rho=\rho_1$ in terms of this number, the critical value
of $m$ being that associated with the largest $\lambda_1\equiv\lambda(\rho_1,\eta;n)$ when $n\in\mathbb{N}$.
This procedure is carried out in Figure~\ref{n_curves} where we consider a
sample of values for $\eta$ (see the caption for details): the horizontal axis records the mode number 
$n/\omega_0$ while the vertical axis shows $\lambda_1$. 
%
%
Strictly speaking, $n\in\mathbb{N}$ but we shall take this parameter to be a positive real
number and notice that the eigenvalue of the problem (\ref{beef_ode},\ref{bc_ode}),
$\omega_0$, will depend on this quantity as well as on $\eta$, i.e., $\omega_0\equiv\omega_0(\eta,n)$.
For $\eta=1$ we find
the curve shown with a continuous line and which consists of two sloping parts separated by a
peak, $P_1$ (corresponding to $n\approx{2.43}$). Henceforth, we shall refer to this
curve as $\mathcal{C}_1$. 
Note that the right-hand part is monotonic decreasing and unbounded but here
only a segment of that curve is shown. The neutral stability curves for the
other values of $\eta=\eta_j>1$ considered are  
$\mathscr{C}_j:=\{(\lambda_1(\rho_1;\eta_j,n),\,n/\omega_0)\;{|}\;n\in\mathbb{R}_{+}\}$
and they all turn out to be part of $\mathscr{C}_1$. The remark made above 
regarding $\mathscr{C}_1$ applies for these curves as well. For the sake of
clarity in Figure~\ref{n_curves} the curves are shifted and shown separately as 
dashed lines, but their top endpoints ($P_2\div P_6$) are marked on $\mathscr{C}_1$ as well.
All such points correspond to the choice $n=1$.\par
The feature illustrated in Figure~\ref{n_curves} is generic and not restricted
to the particular values of $\eta>1$ chosen. A first observation is that the
number of ripples on the compressed side of the block increases with the
non-dimensional thickness $\eta$. When this latter quantity is reasonably large
($\eta>\approx{3}$) the critical mode number given by (\ref{mode_number})
always corresponds to $n=1$. Thus, the behaviour of a very thick block can be 
understood in two different ways: $(i)$ assuming that $n=1$ and
$\eta\gg{1}$ or, conversely, $(ii)$ fixing $\eta=\mathscr{O}(1)$ and letting
$n\gg{1}$. In the former case the {\emph{critical}} mode number will simply be $\pi/\omega(\eta,1)$,
whereas in the latter one the following observation helps: if 
$\eta_1,\,\eta_2>0$ are two given, sufficiently large aspect ratios with $\eta_1<\eta_2$, then
$$
\frac{1}{\omega_0(\eta_2,1)} = \frac{\eta_2}{\omega_0(\eta_1,\eta_2)}\,.
$$
Comparing this with (\ref{mode_number}) assertion $(ii)$ should now be obvious.\par
\begin{figure}[!h]
    \unitlength=1cm
      \psfragscanon
        \psfrag{aa1}[u]{$(a)$}
        \psfrag{aa2}[u]{$(b)$}
        \psfrag{aa3}[u]{$(c)$}
        \psfrag{aa4}[u]{$(d)$}
        \centerline{\includegraphics[width=13cm,height=9.5cm]{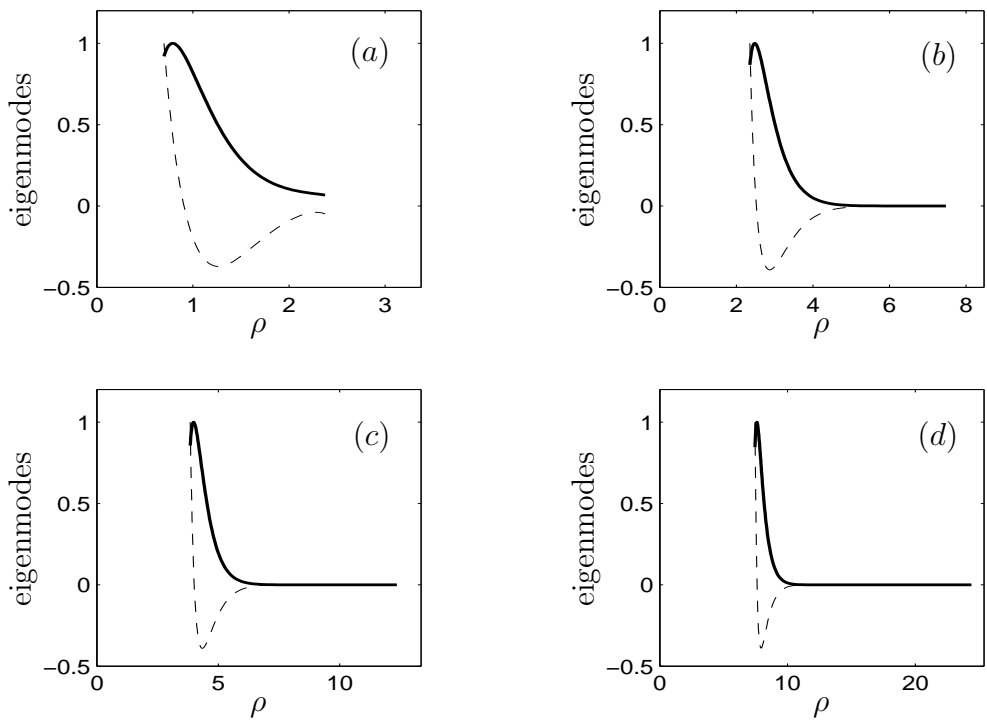}}
        \begin{center}
       \parbox[b]{14cm}
          {\caption[]
           {\small{The eigenfunctions associated with the critical points
          $P_j$ in Figure~\ref{n_curves}: $(a)$~$P_1$ ($\eta=1.0$), $(b)$~$P_2$
          ($\eta=3.0$), $(c)$~$P_3$ ($\eta=5.0$), and $(d)$~$P_4$ ($\eta=10.0$).
          In each plot the continuous line denotes $\rho^{-1}\Phi(\rho)$
          ({\emph{radial displacement}}), while the dashed line is used for
          $\Phi'(\rho)$ ({\emph{azimuthal displacement}}). The range for these
          functions is $\rho_1\leq{\rho}\leq{\rho_2}$ and they are
          suitably normalised so that their maximum amplitude is unity.}}\label{eigenmodes}}
        \end{center}
\end{figure}

It is instructive to gain some insight into the behaviour of the critical
eigenfunctions associated with the $P_j$'s marked on Figure~\ref{n_curves}.
This information is included in Figure~\ref{eigenmodes} where, for the sake of
brevity, we show the radial and azimuthal displacements only for $P_1\div{P_4}$,
as obtained from the two equations in (\ref{potents}). 
%
%
The localisation of the
deformation near the curved inner surface of the bent block when $\eta$
increases is clearly obvious. The stress concentration phenomenon revealed by
these plots is to be expected because the thicker the rubber block, the more difficult is to
bend it, that is, the instability will be likely to occur for small values of the
bending angle. Hence, curvature effects will only be ``felt'' in the immediate
proximity of the bent inner surface. In the remaining of the paper we show that this
behaviour is ideally suited for a singular perturbation analysis.\par
\begin{figure}[!h]
    \unitlength=1cm
      \psfragscanon
        \centerline{\includegraphics[width=10cm,height=10.0cm]{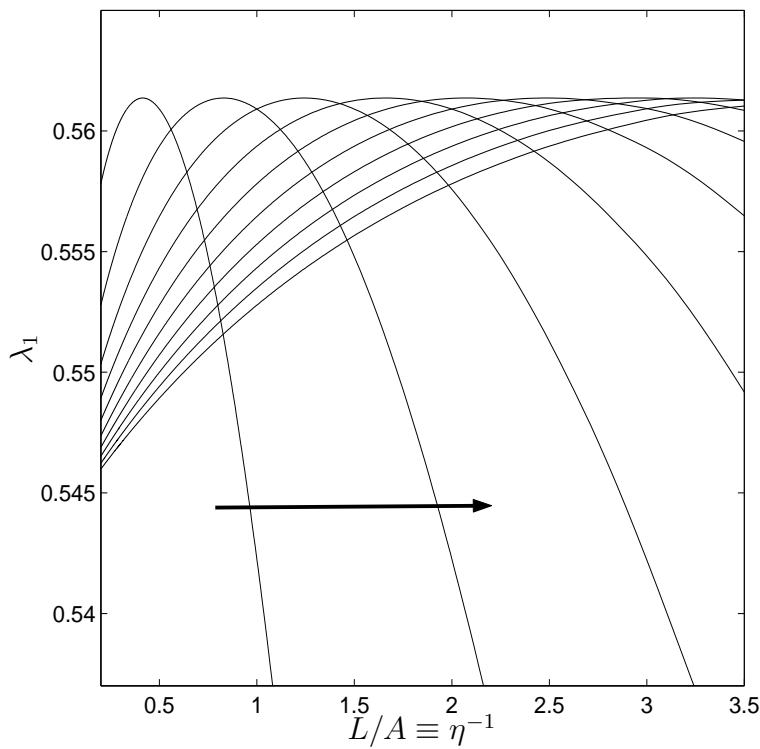}}
        \begin{center}
       \parbox[b]{14cm}
          {\caption[] 
          {\small{ A plot of critical values of $\lambda_1\equiv\lambda(\rho_1)$
          against undeformed length $L/A$ for mode numbers $n=1\div{10}$ (see
          also Figure~$5$ in reference \cite{dmh:99}). The arrow indicates the direction 
          of increasing $n$.}}\label{dmh_pic}}
        \end{center}
\end{figure}

At this juncture some remarks on the method used to identify the critical mode
number are appropriate. At first sight, our work in Figure~\ref{n_curves} 
might appear a little awkward. The coincidence of the curves $\mathscr{C}_j$ 
($j=2,\dots,6$) with $\mathscr{C}_1$ could have been inferred by taking into 
account that the principal stretch $\lambda_1$ is independent of $n$ and depends only on
the product $\omega_0\eta$ -- see (\ref{boundaries}) and (\ref{r_stretch}).
However, we believe that the longer route taken here has the advantage of
clarifying some of the vague statements made by Haughton in \cite{dmh:99}.
He misinterpreted the role played by $n\in\mathbb{N}$ in formula (\ref{mode_number})
and, in Figure~$5$ of his paper, he varied both $n$ and $\eta$ ending up with 
a wrong statement as to the behaviour of the neutral stability curves for the
bending problem. For convenience we reproduce that scenario in our
Figure~\ref{dmh_pic}. Although a different formulation of the eigenproblem
was used in \cite{dmh:99} (without recourse to any
potential function), the results we show are the same (as they should be since 
the height of the rubber block in \cite{dmh:99} was fairly large, $H/A=10$). The only exception
is the unusual feature seen in that paper for $n=1$, which we did not find
with our model. Triantafyllidis \cite{tria:80} seems to have committed a different error by
excluding $\omega_0$ from what he refers to as ``wave-number'' (see Figure~$7$
in his work). That might explain why he did not find an Euler-type buckling instability.  \par
%
%
%
The response curves shown in Figure~\ref{dmh_pic} are reminiscent of similar situations
encountered in the buckling of thin-walled structures (e.g.,
\cite{alfutov:00}). In that particular type of situation $n$ represents the number
of half-waves of the instability pattern and plots like the one
shown above can be used to infer the wavelength of the buckling pattern from
knowledge of some aspect ratio (related recent work on thin-film instabilities
can be found in \cite{cdc_dmh_1,cdc_apb:2007a,cdc_apb:2007b,cdc_apb:2007c}).
However, in the present context such extrapolations appear to provide
misleading information for obvious reasons. Also, in the limit $A/L\to\infty$
the critical principal stretch would have to
be equal to the value $0.544$ predicted by Biot's analysis for a neo-Hookean half-plane
in compression (cf. \cite{gent:99,biot:65}). This is clearly not the case in Figure~\ref{dmh_pic} but,
on the other hand, the earlier results of Figure~\ref{n_curves} do anticipate
this expectation.
%
%
%
\section{Stress concentration for $n\gg{1}$}
%
\noindent The mathematical structure of the boundary-value problem derived in
\S2 is akin to a number of situations investigated recently in the literature 
by Fu {\emph{et al.}} \cite{fu:02a,fu:02b} and Haughton \& Chen \cite{dmh_chen:02}. 
Broadly speaking, these authors encountered a 
particular occurrence of turning points (e.g., \cite{bender:99})
or repeated roots in the characteristic equations associated with
bifurcation analyses for everted cylindrical/spherical shells. It was
stated that such special points could aid in detecting sites of high-stress
concentration within elastic solids.
Furthermore, in light of the recent work on edge-buckling of thin 
films \cite{cdc_dmh_1,cdc_apb:2007a,cdc_apb:2007b,cdc_apb:2007c}, 
it would appear very reasonable to conclude that Fu's observation
might go a long way towards explaining the localised behaviour 
seen in Figure~\ref{eigenmodes}. That this is not true we are
going to see in \S\ref{blt}, but before we pursue those issues
it is important to gain an understanding of the relevance of WKB techniques in the present
context. According to the previous interpretation of the parameters $\eta\equiv{A}/L$ and
$n$ (defined in equation (\ref{mode_number})), the bifurcation of a thick
rubber block ($\eta\gg{1}$) can be understood by taking $\eta={1}$ and allowing
$n\gg{1}$. This is precisely what we do in the remainder of the paper.
%
\subsection{WKB approach}
\label{wkb_approach}
%
%
\noindent The WKB method is a simple and efficient tool for dealing with
variable-coefficient linear differential equations containing certain small or
large parameters. We shall exploit the presence of $\mu\equiv\pi{n}\gg{1}$ in our
eigenproblem to describe the dependence of $\omega_0$ (or $\lambda_1$) on this large parameter.\par
A WKB solution of (\ref{beef_ode}) is sought in the form
\begin{subequations}
\label{wkb_ansatz}
\begin{alignat}{1}
\label{wkb_ansatz:1}
\Phi(\rho) &= Y(\rho)\exp\left(\mu\int_{\rho_1}^{\rho} S(\xi)\,\mathrm{d}\xi\right)\,,\\[0.2cm]
\label{wkb_ansatz:2}
Y(\rho) &= Y_0(\rho)+\frac{1}{\mu}Y_1(\rho)+\frac{1}{\mu^2}Y_2(\rho)+\dots
\end{alignat}
\end{subequations}
where $S\equiv{S}(\rho)$ is one of the roots of the characteristic equation
$$
S^4 - \left(\omega_0^2\rho^2+\frac{1}{\omega_0^2\rho^2}\right)S^2 + 1 = 0\,,
$$
and $Y_j(\rho)$ ($j=0,1,\dots$) are functions that are to be determined
sequentially by substituting the ansatz (\ref{wkb_ansatz}) into
(\ref{beef_ode}), and then solving the differential equations obtained by
setting to zero the coefficients of like powers of $\mu$. The above bi-quadratic
has four real roots that will be labelled
$$
S_1^{(\pm)}(\rho):= \pm\omega_0\rho,\qquad S_2^{(\pm)}(\rho):= \pm\frac{1}{\omega_0\rho}\,,
$$
and they lead to a set of linearly independent (approximate) solutions for (\ref{beef_ode}). Given our
experience with the direct numerical simulations of \S\ref{numerics}, it
is expected that only the exponentials corresponding to $S_{1,2}^{(-)}(\rho)$ need to
be used in order to capture the through-thickness localised behaviour. The
superscripts ``$1$" and ``$2$" will be used to identify quantities     
associated with these characteristic exponents in (\ref{wkb_ansatz}).\par
The determinantal equation that follows by imposing the boundary conditions (\ref{bc_ode})
at $\rho=\rho_{1}$ on the WKB solutions $\Phi^{(1)}$ and $\Phi^{(2)}$ has the
form
\begin{equation}
\label{det}
U_1(\rho_1)V_2(\rho_1)-U_2(\rho_1)V_1(\rho_1) = 0\,,
\end{equation}
where
\begin{alignat}{1}
U_j(\rho) &:= \Phi^{(j)'''} - \mu^2\left(\omega_0^2\rho^2+\frac{2}{\omega_0^2\rho^2}\right)\Phi^{(j)'} 
        + \frac{\mu^2}{\rho}\left(\omega_0^2\rho^2+\frac{2}{\omega_0^2\rho^2}\right)\Phi^{(j)}\,,\\[0.2cm]
V_j(\rho) &:= \Phi^{(j)''} -\frac{1}{\rho}\Phi^{(j)'}+\frac{\mu^2}{\omega_0^2\rho^2}\Phi^{(j)}\,,\qquad j=1,\,2\,.
\end{alignat}
When calculating $\Phi^{(j)}$ ($j=1,2$) we shall ignore terms of order $\mathscr{O}(\mu^{-2})$ and
higher in the ansatz (\ref{wkb_ansatz:2}). These solutions are fixed by
routinely solving a series of non-homogeneous linear differential equations. 
The various multiplicative and additive constants in the expressions of those
functions can be chosen (without loss of generality) to be unity or, respectively, equal to
zero. The final results are
\begin{alignat*}{2}
Y_0^{(1)}(\rho) & = \frac{\rho}{(1-\omega_0^4\rho^4)^{1/2}}\,,\qquad\quad
  Y_1^{(1)}(\rho) &&= -\frac{5\omega_0^8\rho^8+10\omega_0^4\rho^4-3}
                            {4\omega_0\rho^2(1-\omega_0^4\rho^4)^2}\,Y_0^{(1)}(\rho)\,,\\[0.2cm]
Y_0^{(2)}(\rho) & = \frac{\rho^2}{(1-\omega_0^4\rho^4)^{1/2}}\,,\qquad\quad
  Y_1^{(2)}(\rho) &&= \frac{3\omega_0(\omega_0^4\rho^4+1)}{2(1-\omega_0^4\rho^4)^2}\,Y_0^{(2)}(\rho)\,,
\end{alignat*} 
and thus,
\begin{equation}
\label{wkb2term}
\Phi^{(j)}(\rho)\approx\left\{Y_0^{(j)}(\rho)+\frac{1}{\mu}Y_1^{(j)}(\rho)\right\}
               \exp\left(\mu\int_{\rho_1}^{\rho} S_j^{(-)}(\xi)\,\mathrm{d}\xi\right)\,,\qquad j=1,\,2\,.
\end{equation}
It must be noted that $Y_{j}^{(j+1)}$ ($j=0,1$)  blow up when
$\rho=\overline{\rho}\equiv\omega_0^{-1}\in(\rho_1,\,\rho_2)$; in the language of differential
equations this represents a (multiple) turning point of the differential
equation (\ref{beef_ode}). Such
points are usually defined as those values of the independent variable for
which some of the roots of the characteristic equation coalesce. For this
particular example, both  
$S_1^{(+)}(\overline{\rho}) = S_2^{(+)}(\overline{\rho})$ 
and 
$S_1^{(-)}(\overline{\rho}) = S_2^{(-)}(\overline{\rho})$, {\emph{i.e.}},
two pairs of roots merge. Strictly speaking, the validity of the above formulae 
for $Y_j^{(j+1)}$ ($j=0,1$) requires 
$|\rho-\overline{\rho}|\gg\mu^{-1/2}$. 
Although $\overline{\rho}$ depends on the unknown
eigenvalue, our numerical experiments suggest that the turning point remains
confined to the central part of the interval
$(\rho_1,\,\rho_2)$.\par
On making use of (\ref{wkb2term}) into (\ref{det}), we are able to expand
the determinantal equation in decreasing integral powers of $\mu\gg{1}$,
\begin{equation}
\label{det_expanded}
\Gamma_0(\lambda_1)+\Gamma_1(\omega_0,\lambda_1)\frac{1}{\mu}+
        \Gamma_2(\omega_0,\lambda_1)\frac{1}{\mu^2} + \dots = 0\,,
\end{equation}
with
\begin{alignat*}{1}
\Gamma_0(z) &:= 8z^2(z^2+1)^2(z^3-z^2+z+1)(z^3+z^2+z-1)\,,\\[0.2cm]
\Gamma_1(\omega_0,z) &:= 2\omega_0(4z^{12}+15z^{10}+23z^8+8z^4-7z^2-3)\,,
\end{alignat*}
and
{\setlength{\multlinegap}{30pt}
\begin{multline}
\Gamma_2(\omega_0,z):=\frac{\omega_0^2}{(z^2+1)^2(z^2-1)^4}\big{(}6z^{22}-18z^{20}-37z^{18}-477z^{16}-1118z^{14}\\
          -930z^{12}-600z^{10}-4z^8-120z^6-140z^4+45z^2+33\big{)}\,.\nonumber
\end{multline}}
\hspace*{-0.2cm}The solution of (\ref{det_expanded}) yields approximations for both the
critical bending angle $\omega_0$, and the principal stretch $\lambda_1$.
For the sake of brevity we record only the final results here
\begin{gather}
\label{omega0}
\omega_0 = \overline{\Omega}_0 + \frac{\overline{\Omega}_1}{\mu} + \frac{\overline{\Omega}_2}{\mu^2}+\dots\,,\\[0.2cm]
\overline{\Omega}_0=0.771844\,,\qquad
\overline{\Omega}_1=-1.305565\,,\qquad
\overline{\Omega}_2=15.39664\,,\nonumber
\end{gather}
and 
\begin{gather}
\label{lambda1}
\lambda_1 = \Lambda_0 + \frac{\Lambda_1}{\mu} +\frac{\Lambda_2}{\mu^2}+\dots\,,\\[0.2cm]
\Lambda_0=0.543689\,,\qquad
\Lambda_1=0.385922\,,\qquad
\Lambda_2=-4.184333\,.\nonumber
\end{gather}
As one would expect, $\Lambda_0\approx{0.544}$ represents the critical value of the principal
stretch for the surface instability of a compressed neo-Hookean half-space
(cf. \cite{biot:65}); the next-order corrections in formula (\ref{lambda1})
account for the finite size of the rubber block. To assess the usefulness of 
the two asymptotic results (\ref{omega0}) and (\ref{lambda1}),
a set of comparisons with direct numerical simulations is recorded in 
Table~\ref{wkb_compare}. The agreement is excellent for both $\omega_0$ and 
$\lambda_1$; in particular, we find that the relative accuracy (RA) associated
with $\omega_0$ ranges between $1.4\%$ ($n=7$) and $0.8\%$ ($n=20$).
The approximation of $\lambda_1$ is even better, for RA is at most $0.4\%$
($n=7$) in all cases considered. \par
The WKB analysis laid out above has the advantage of producing a
robust approximation for $\omega_0$ (or $\lambda_1$) with minimum effort. The presence of the
turning point, however, is worrying because it tends to obscure the true
nature of the localised behaviour exhibited by (\ref{beef_ode}). It is not immediately
clear whether such behaviour has anything to do with the turning
point and, thus, a change of tack is imperative. It will shortly become obvious
that conventional boundary-layer techniques are better suited for understanding the underlying
mathematical structure responsible for the scenario depicted in
Figure~\ref{eigenmodes}. The details of that particular approach are highlighted next.
\begin{table}[!h]
\begin{center}
\parbox[b]{14.0cm}
{\caption{\small{Comparisons between direct numerical simulations of the
          eigenproblem (\ref{beef_ode},\ref{bc_ode})
          and the asymptotic results recorded in the formulae (\ref{omega0}) and
          (\ref{lambda1}). The bending angle and the azimuthal stretch on the
          inner boundary associated with the former set of values are denoted
          by $\omega^{\text{num}}$ and, respectively, $\lambda_1^{\text{num}}$. The corresponding
          asymptotic quantities are identified as $\omega_0^{\text{asy}}$ 
          and $\lambda_1^{\text{asy}}$.}}\label{wkb_compare}}\\[0.3cm]
\begin{tabularx}{110mm}{cXXXX}
\hline\\[-0.3cm]
$n$ & $\omega_0^{\text{num}}$ & $\omega_0^{\text{asy}}$ & $\lambda_1^{\text{num}}$ & $\lambda_1^{\text{asy}}$ \\[0.1cm]
\hline
7  &  0.744313103   &   0.733652725  &     0.555301084  &    0.552585690\\
8  &  0.744272309   &   0.736778030  &     0.554325988  &    0.552419947\\
9  &  0.744928414   &   0.739453964  &     0.553494699  &    0.552104104\\
10 &  0.745886633   &   0.741762924  &     0.552780068  &    0.551733662\\
11 &  0.746957132   &   0.743772049  &     0.552160229  &    0.551352710\\
12 &  0.748046186   &   0.745533828  &     0.551618218  &    0.550981720\\
13 &  0.749107553   &   0.747090530  &     0.551140475  &    0.550629796\\
14 &  0.750119338   &   0.748474960  &     0.550716526  &    0.550300415\\
15 &  0.751072409   &   0.749714126  &     0.550337796  &    0.549994245\\
16 &  0.751964384   &   0.750829238  &     0.549997574  &    0.549710574\\
17 &  0.752796399   &   0.751837997  &     0.549690282  &    0.549448050\\
18 &  0.753571374   &   0.752754745  &     0.549411416  &    0.549205075\\
19 &  0.754293016   &   0.753591537  &     0.549157200  &    0.548980000\\
20 &  0.754965302   &   0.754358139  &     0.548924583  &    0.548771235\\
\hline
\end{tabularx}
\end{center}
\end{table}
%
%
%
\subsection{Boundary-layer analysis}
\label{blt}
%
%
\noindent To begin, we introduce the stretched variable $X=\mathscr{O}(1)$ such that
$\rho=\rho_1+X\mu^{-1}$, and look for solutions of (\ref{beef_ode}) with
\begin{subequations}
\label{blayer_antz}
\begin{alignat}{1}
\label{blayer_antz:1}
W(X) &= W_0(X)+W_1(X)\frac{1}{\mu}+W_2(X)\frac{1}{\mu^2}+\dots\,,\\[0.2cm]
\label{blayer_antz:2}
\omega_0 &= \Omega_0+\frac{\Omega_1}{\mu}+\frac{\Omega_2}{\mu^2}+\dots\,,\\[0.2cm]
\label{blayer_antz:3}
\rho_1 &= \Delta_0+\frac{\Delta_1}{\mu}+\frac{\Delta_2}{\mu^2}+\dots\,.
\end{alignat}
\end{subequations}
Although (\ref{blayer_antz:2}) and (\ref{blayer_antz:3}) are not independent,
it helps to expand $\rho_1$ in the form suggested here. Of course, when
solving the governing equations for the coefficients $W_j(X)$ ($j=0,1,\dots$)
one has to remember formula (\ref{boundaries}) and replace the $\Delta_k$'s with their
expressions in terms of the $\Omega_j$ ($j=0,1,\dots,k$).\par
On substituting (\ref{blayer_antz}) into (\ref{beef_ode}) we find a hierarchy
of differential equations
\begin{equation}
\label{blayer_ode}
\mathcal{L}_{\text{BL}}[W_k] =
\sum_{i=0}^{k-1}\sum_{j=1}^{3}A_{ij}^{(k)}\frac{d^jW_i}{dX^j}\qquad\quad
(k\geq{0})\,,
\end{equation}
in which
$$
\mathcal{L}_{\text{BL}}:=\frac{d^4}{dX^4}-\left(\zeta_0^2+\frac{1}{\zeta_0^2}\right)\frac{d^2}{dX^2} + 1
$$
is the boundary-layer (BL) differential operator and $\zeta_0:=\Delta_0\Omega_0$. The quantities
$A_{ij}^{(k)}\equiv A_{ij}^{(k)}(X)$ will be defined as
we go along, with the convention that $A_{ij}^{(0)}\equiv{0}$. In contrast to
the WKB analysis the general solution of each one of the equations in
(\ref{blayer_ode}) is trivially found, for $\mathcal{L}_{\text{BL}}$ has
constant coefficients.\par
The equations (\ref{blayer_ode}) are solved subject to two types of boundary
conditions. The first set is obtained from (\ref{bc_ode}) with the help of
the ansatz (\ref{blayer_antz}), and can be cast in the general form
\begin{subequations}
\label{blayer_bc}
\begin{alignat}{2}
\label{blayer_bc:1}
\mathcal{H}_1[W_k] &=\sum_{i=0}^{k-1}\sum_{j=0,1}B_{ij}^{(k)}\frac{d^jW_i}{dX^j}\,,\qquad 
                                          &&\text{for}\;X=0\,,  \\[0.2cm]
\label{blayer_bc:2}
\mathcal{H}_2[W_k] &= \sum_{i=0}^{k-1}\sum_{j=0,1}C_{ij}^{(k)}\frac{d^jW_i}{dX^j}\,,\qquad
                                          &&\text{for}\;X=0\,,
\end{alignat}
\end{subequations}
where
$$
\mathcal{H}_1:=\frac{d^3}{dX^3}-\left(\zeta_0^2+\frac{2}{\zeta_0^2}\right)\frac{d}{dX}
\qquad\text{and}\qquad
\mathcal{H}_2:=\frac{d^2}{dX^2}+\frac{1}{\zeta_0^2}\,;
$$
the remark made for the $A_{ij}^{(k)}$\,'s applies to the boundary coefficients
$B_{ij}^{(k)}$ and $C_{ij}^{(k)}$ as well.\par
The second set of boundary
conditions is motivated by the numerical experiments illustrated in
Figure~\ref{eigenmodes} and involves the requirement that
\begin{equation}
\label{decay}
\frac{d^jW_i}{dX^j}\to{0}\qquad \text{as}\quad X\to\infty\,,
\end{equation}
for $i\geq{0}$, $j=0,\dots,3$. It might also be worth pointing out that the 
solution of equation (\ref{beef_ode}) is exponentially 
small in the outer layer (cf.\S\ref{wkb_approach}), so that
(\ref{decay}) can be viewed as matching conditions between the inner and the
outer solutions.\par
The leading order problem for $W_0(X)$ is homogeneous and consists of the
differential equation (\ref{blayer_ode}) for $k=0$, together with the
boundary conditions (\ref{blayer_bc}). Rejecting the exponentially 
growing contributions and imposing the normalisation condition $W_0(X=0)=1$,
it follows that
\begin{equation}
W_0(X)=\frac{2}{1-\zeta_0^4}\exp(-\zeta_0{X})-\frac{1+\zeta_0^4}{1-\zeta_0^4}\exp(-X/\zeta_0)\,.
\end{equation}
When this function is substituted into the boundary conditions, one obtains an
algebraic equation,
$$
\zeta_0^8+2\zeta_0^4-4\zeta_0^2+1=0\,,
$$
whose only acceptable solution is $\zeta_0\approx 0.543689$. The result is identical
to $\Lambda_0$ found in \S\ref{wkb_approach} and the same turns out to be
true for $\Omega_0$ in (\ref{blayer_antz:2}).\par 
The next order problem corresponds to taking $k=1$ in (\ref{blayer_ode}) and
(\ref{blayer_bc}). The coefficients that appear in these equations 
are 
\begin{alignat*}{2}
A_{01}^{(1)} &:= \frac{1}{\Delta_0\zeta_0^2}(\zeta_0^4-3)\,,\quad
&&A_{02}^{(1)} := \frac{2}{\zeta_0^3}(\Delta_0\Omega_1+\Delta_1\Omega_0+\Omega_0{X})(\zeta_0^4-1)\,,\quad
A_{03}^{(1)} := \frac{2}{\Delta_0}\,,\\[0.2cm]
B_{00}^{(1)} &:= -\frac{\Omega_0}{\zeta_0^3}(\zeta_0^4+2)\,,\quad
&&B_{01}^{(1)} := \frac{2}{\zeta_0^3}(\Delta_0\Omega_1+\Delta_1\Omega_0)(\zeta_0^4-2)\,,\\[0.2cm]
C_{00}^{(1)} &:=\frac{2}{\zeta_0^3}(\Delta_0\Omega_1+\Delta_1\Omega_0)\,,\quad
&&C_{01}^{(1)} :=\frac{1}{2}A_{03}^{(1)}\,.
\end{alignat*}
The first-order correction $\Omega_1$ in the expansion (\ref{blayer_antz:2}) of the eigenvalue
$\omega_0$ is recovered by enforcing the Fredholm solvability condition
on the {\emph{non-homogeneous}} problem satisfied by $W_1(X)$. The task is
simplified by the observation that the homogeneous problem for $W_0(X)$ is self-adjoint.
Standard calculations show that this constraint amounts to
{\setlength{\multlinegap}{45pt}
\begin{multline}
\label{fredholm}
C_{01}^{(1)}\left\{\frac{dW_0}{dX}(0)\right\}^2+\left\{C_{00}^{(1)}-B_{01}^{(1)}\right\}W_0(0)\frac{dW_0}{dX}(0)
                     -B_{00}^{(1)}\left\{W_0(0)\right\}^2 \\ 
=\int_0^{\infty}W_0(\xi)\sum_{j=1}^{3}A_{0j}^{(1)}\frac{d^jW_0}{dX^j}(\xi)\,\mathrm{d}\xi\,.
\end{multline}}
\hspace*{-0.1cm}Notice that the integral on the right-hand side in (\ref{fredholm}) is evaluated
analytically, and thus the solvability condition will reduce to a linear
equation in $\Omega_1$. The solution $\Omega_1\approx-1.305562$ is, for
all practical purposes, identical to the WKB result obtained earlier.\par
The pattern of the boundary-layer approach for the pure bending problem
is now clear: at each step one will have to impose a solvability condition 
for finding $\Omega_j$ that features in (\ref{blayer_antz:2}), and then solve (exactly) a
non-homogeneous fourth-order boundary-value problem in order to get $W_j(X)$.
The algebraic manipulations become increasingly unwieldy as we move to
further orders, but symbolic algebra packages help considerably. We 
have imposed the solvability condition for the $W_2$-problem and found that
the value of $\Omega_2$ predicted agrees with $\overline{\Omega}_2$ to
within five significant digits; for completeness, the coefficients needed to set up that problem
are recorded below  
\begin{alignat*}{1}
A_{11}^{(2)} &:= \frac{\Omega_0}{\zeta_0^3}\left(\zeta_0^4-3\right)\,,\quad
A_{12}^{(2)} := A_{02}^{(1)}\,,\quad
A_{13}^{(2)} := A_{03}^{(1)}\,,\\[0.2cm]
A_{01}^{(2)} &:= \frac{1}{\zeta_0^4}\left[\Omega_0^2(X+\Delta_1)(\zeta_0^4+9)+
                2\Omega_1(\zeta_0^4+3)\zeta_0\right]\,,\quad
A_{03}^{(2)} := -\frac{2}{\Delta_0^2}(X+\Delta_1)\,,\\[0.2cm]
A_{02}^{(2)} &:=
\frac{1}{\zeta_0^4}\bigg{\{}2\Delta_0\Omega_0(\Delta_0\Omega_2+\Delta_2\Omega_0)(\zeta_0^4-1)
             +\left[\Omega_0^2(X+\Delta_1)^2+\Delta_0^2\Omega_1^2\right](\zeta_0^4+3)\\[0.2cm]
          &\hspace{8.5cm}   +4\Omega_1(X+\Delta_1)(\zeta_0^4+1)\zeta_0-3\Omega_0^2\zeta_0^2\bigg{\}}\,,\\[0.2cm]
B_{00}^{(2)} &:=-\frac{1}{\zeta_0^4}\left[\Delta_1\Omega_0^2(\zeta_0^4-6)+2\Omega_1(\zeta_0^4-2)\zeta_0\right]\,,\\[0.2cm]      
B_{01}^{(2)}&:=\frac{1}{\zeta_0^4}\left[2(\Delta_0\Omega_2+\Delta_2\Omega_0)(\zeta_0^4-2)\zeta_0
              +(\Delta_0^2\Omega_1^2+\Delta_1^2\Omega_0^2)(\zeta_0^4+6)+4\Delta_1\Omega_1(\zeta_0^4+2)\zeta_0\right]\,,\\[0.2cm]
B_{10}^{(2)} &:=B_{00}^{(1)}\,,\quad 
B_{11}^{(2)} := B_{01}^{(1)}\,.\\[0.2cm]
C_{00}^{(2)} &=-\frac{1}{\zeta_0^4}\left[3(\Delta_0^2\Omega_1^2+\Delta_1^2\Omega_0^2)
                    - 2\zeta_0(\Omega_0\Delta_2+\Delta_0\Omega_2-2\Delta_1\Omega_1)\right]\,,\\[0.2cm]
C_{01}^{(2)} &= -\frac{\Delta_1}{\Delta_0^2}\,,\quad 
C_{10}^{(2)} = C_{00}^{(1)}\,,\quad
C_{11}^{(2)} = C_{01}^{(1)}.
\end{alignat*}
It might appear that our boundary-layer approach is a by-product of adopting
the simple neo-Hookean form for the constitutive response of the bulk
material. However, this impression is only apparent, for choosing a
strain-energy function of the form
$$
W(\lambda_r,\lambda_{\theta})\propto (\lambda_r^q +
\lambda_{\theta}^q)\,,\qquad (q>0)\,,
$$ 
the same mathematical structure persists. In this case the interpretation
of the asymptotic results become more difficult because now loss of
elliptiticy will be encountered for some value of $q\neq{2}$; such issues will
be discussed elsewhere \cite{cdc_md}.
%
%
%
\section{Concluding remarks}
%
%
%
\noindent We have re-examined the bifurcations in cylindrical bending
of a thick rubber block under the assumption of plane-strain deformation. The
constitutive behaviour was taken to be that of a neo-Hookean incompressible
solid, the reason for this being twofold: we wanted to $(a)$ exclude any 
material instabilities and, $(b)$ simplify our equations as much as possible.
The outcome turned out to be a simple fourth-order eigenproblem with variable
coefficients. Direct numerical simulations
and singular perturbation methods were employed to unravel the origins of the
rippling pattern triggered on the compressed face of the block, when the
bending angle is sufficiently large.
It has been shown in \S\ref{numerics} that previous investigators
\cite{tria:80,dmh:99} misinterpreted the definition of the so-called
``{\emph{mode number}}'' and thus made several erroneous statements.
In particular, we want to re-iterate here that blocks of a large but {\emph{finite thickness}}
will always experience an Euler-type buckling instability with a well defined number of
ripples. It is only in the limit of an {\emph{infinitely thick}} block that 
one finds the degenerate surface instability. Our results deal with
the neo-Hookean material, but it is believed that the above statement remains
valid in other cases as well. It seems quite unlikely that the choice of
constitutive law would have any dramatic repercussions on the overall arguments
presented in this paper. \par
Our work has also demonstrated that the use of WKB methods
in the context of incremental elasticity is unnecessary and even misleading. 
The multiple turning point featuring in the present eigenproblem has little to
do with the tendency of the rippling deformation to confine itself near the
inner curved surface of the block. 
Interestingly though, a simple-minded boundary-layer analysis was able to expose the
nature of the localisation in a trivial way. In spite of the original
complexity of the problem, we found that if the rubber block is sufficiently
thick, its possible bifurcations from the cylindrical configuration are
governed by constant-coefficient differential equations -- easily solvable in closed
form. This has important overall implications since preliminary calculations
indicate that the problems taken up in \cite{fu:02a,fu:02b,dmh_chen:02}) are
amenable to a similar boundary-layer analysis. We shall report the
corresponding details in a forthcoming study \cite{cdc_md}.
%
%
%

%

%
%
%
%
\end{document}